# Exceptional features in nonlinear Hermitian systems

Liang Fang[1, #], Kai Bai[1, #], Cheng Guo[2], Tian-Rui Liu[1], Jia-Zheng Li[1], and Meng Xiao[1, 3, †]

[1]Key Laboratory of Artificial Micro- and Nano-structures of Ministry of Education and School of Physics and Technology, Wuhan University, Wuhan, China

[2] Ginzton Laboratory and Department of Electrical Engineering, Stanford University, Stanford, California 94305, USA

[3]Wuhan Institute of Quantum Technology, Wuhan 430206, China

[#]Co-first author

Corresponding E-mail: [†]phmxiao@whu.edu.cn

## Abstract

Non-Hermitian systems and their topological singularities, such as exceptional points (EPs), lines, and surfaces, have recently attracted intense interest. The investigation of these exceptional constituents has led to fruitful applications. The responsivity of the eigenvalue diverges at EPs, and chiral state transfer occurs when encircling an EP. Traditionally, it was believed that these exceptional features were unique to non-Hermitian systems requiring gain, loss, or nonreciprocal hopping. Here, we show that these exceptional features are also present in nonlinear Hermitian systems. We consider two coupled resonators with Kerr nonlinearity in one resonator, and no non-Hermitian terms. We identify EP-like points (ELPs) on the eigenspectra where the critical behaviors are the same as those of typical EPs. Additionally, this nonlinear Hermitian system can be mapped to linear non-Hermitian systems, with ELPs corresponding to EPs. We also demonstrate that encirclement around an ELP in the parameter space leads to unique chiral state transfer behavior.

**Introduction**.

The last two decades have witnessed the prosperity of non-Hermitian effects across various research frontiers of physics [1–10]. Bender and Boettcher's milestone work reveals that a specific class of complex Hamiltonians can manifest a wholly real spectrum, provided that parity-time reversal (PT) symmetry remains intact [1]. One particularly intuitive example of such a system involves two coupled resonators, each featuring gain and loss, as sketched in Fig. 1(a). When the ratio between gain/loss and the coupling ($g/\kappa$) is small, the spectrum remains real; while when $g/\kappa$ is large, gain and loss are not balanced inside each eigenmode and the spectra become complex [Fig. 1(b)]. The phase transition happens at $g/\kappa = 1$ where the system Hamiltonian becomes defective, and such a point is dubbed an exceptional point (EP). Despite its simplicity, this Hamiltonian has become a fertile ground for novel concepts and fruitful applications [11–20]. For instance, encircling the EP in the parameter space can encounter a nonadiabatic transition, which further leads to captivating chiral behavior where the circling direction determines the final state [17–19].

Incorporating nonlinear effects into non-Hermitian systems introduces novel concepts and yields intriguing outcomes [21–26]. For instance, nonlinear Kerr effects can facilitate the transition of a system between PT-symmetric and PT-broken regions or vice versa [24]. Besides the typical Hermitian nonlinear effects such as the Kerr effect, nonlinear effects are also natural for non-Hermitian systems when considering the dependence of gain and loss coefficients on the field amplitude. Figure 1(c) shows two coupled resonators with a saturable gain inside one of the resonators. Over a proper parameter range, such a system maintains PT-symmetric automatically, which further enables robust wireless power transfer [25,26]. In addition, the nonlinear saturable gain herein also introduces an auxiliary unstable mode and enables higher-order exceptional points even within such a two-resonator system [27]. Figure 1(d) shows the typical band structure as a function of the detuning between two resonators. Inside the bi-stable region (light gray region), the system exhibits two stable modes (blue) and one unstable mode (red). These three modes together form a S-shape curve. At the boundary of the bi-stable region, two nonlinear EPs (NEPs) are formed when the unstable mode collapses with the stable modes. Beyond the bi-stable region, the band structure becomes complex as in typical linear non-Hermitian systems, e.g., Fig. 1(b). The critical behavior of the eigenspectra when approaching the NEPs is the same as that for EPs in linear systems [27–29].

The S-shape curve with two stable modes and one unstable mode is common in Hermitian nonlinear systems when considering the Kerr effect. Figure 1(e) sketches one such model within a two-resonator system. The resonance frequency of one of the resonators depends on the wave amplitude inside that resonator through $\omega_A + \alpha|\psi_1|^2$, where $\omega_A$ is the resonance frequency at low field limit and $\alpha$ is the Kerr coefficient. Figure 1(f) shows the typical eigenmode spectrum of such a system as a function of $\alpha$. Similar to Fig. 1(d), we can observe a S-shape curve and a bi-stable region (light gray region). Beyond the bi-stable region, the eigenfrequencies of the two modes become complex while the remaining one is still real. The transitional points between real and complex spectra are analogous to EPs and thus dubbed EP-like points (ELPs). Considering the similar features presented in these systems, our work aims to systematically analyze the phenomena of Hermitian nonlinear systems that are akin to non-Hermitian systems. We show that such a nonlinear system can be mapped to higher-dimensional linear systems. Although the mapping above is not isomorphic (i.e., one nonlinear system can be mapped to an infinite number of higher-dimensional linear systems), all these linear systems must be non-Hermitian under iso-spectral transformation [30]. In addition, all these linear systems share the same Sylvester matrix with the ELPs mapped to linear EPs. In other words, the order of EPs and thus the critical behavior when approaching these EPs are preserved under these mappings. We have also demonstrated the unique chiral behavior in nonadiabatic transition when encircling the ELPs in the parameter space.

**Model.**

As sketched in Fig. 1(e), we consider two coupled resonators, and the resonance frequency of resonator A exhibits a Kerr form dependence on the wave amplitude inside that resonator. The system dynamics follow the nonlinear Schrödinger equation

$$i\frac{d}{dt}\begin{pmatrix}\psi_A\\\psi_B\end{pmatrix} = H_{NL}\begin{pmatrix}\psi_A\\\psi_B\end{pmatrix} = \begin{pmatrix}\omega_A+\alpha|\psi_A|^2 & \kappa\\\kappa & \omega_B\end{pmatrix}\begin{pmatrix}\psi_A\\\psi_B\end{pmatrix},\tag{1}$$

where $H_{NL}$ is the nonlinear Hamiltonian, $\alpha$ is the Kerr coefficient, $\kappa$ is the coupling strength, and $\psi_A$ ($\psi_B$) and $\omega_A$ ($\omega_B$) are the wave amplitude and resonance frequency inside the resonator A (B), respectively. We make no distinction between electrons or photons in this work, and for simplicity, we set the constants such as elementary charge $e$ and reduced Planck constant $\hbar$ to 1. Thus, all the parameters inside Eq. (1) are appropriately chosen as unitless throughout this letter. To maintain consistency, we set

$$|\psi_A|^2 + |\psi_B|^2 = 1.\tag{2}$$

This equation corresponds to the conservation of probability and energy for matter and classical waves, respectively. For classical waves such as light, the total intensity varies for different systems and different initial conditions, and the variation of intensity can be taken into consideration by varying the coefficient $\alpha$. As a global frequency shift does not change the underlying physics, we set $\omega_B = 0$ and define the detuning as $\Delta_\omega \equiv \omega_A - \omega_B$. Meanwhile, we set $\kappa$ to be real, either positive or negative. The behavior of systems with complex $\kappa$ can be obtained by multiplying $\psi_A$ with the phase factor of $\kappa$. [See details in the Supplementary Materials (SM) Sec. I.]

**Eigenspectra of the nonlinear Hamiltonian.**

We start with the eigenequation of $H_{NL}$,

$$H_{NL}\Psi_n = \Omega_n\Psi_n,\tag{3}$$

where $\Omega_n$ and $\Psi_n = \left(\psi_{n,A}, \psi_{n,B}\right)^T$ are the $n$-th eigenfrequency and eigenstate of the system, respectively. The eigenfrequency satisfies the characteristic equation

$$P(\Omega) = \Omega^4 - \Omega^3(\Delta_\omega + \alpha) - \Omega\kappa^2\Delta_\omega - \kappa^4 = 0,\tag{4}$$

where the band label subscript is omitted for simplicity. $P(\Omega)$ is a fourth-order polynomial of $\Omega$ and thus four eigenfrequencies can be identified from Eq. (4). Since all the other coefficients in Eq. (4) are purely real, the four eigenfrequencies are either real or come in complex conjugate pairs. Figure 2(a) shows the eigenfrequencies of the system as a function of Kerr coefficient $\alpha$ and detuning $\Delta_\omega$ for fixed coupling strength $\kappa$. Here we only plot the purely real eigenvalues. For the parameter space we consider, the first band (light gray) is purely real and isolated from other higher bands; other bands are real only over some part of the parameter space. When $\alpha$ is large enough ($\alpha > \alpha_{EX} = 4$), we can see the typical S-shape curve formed among higher bands as we vary $\Delta_\omega$. Here, the eigenstates on band 2 and band 4 (pink and red surfaces) are stable in dynamics, while those on band 3 (blue surface) are unstable.

We proceed to show both the real and complex parts of the solution in Eq. (4) along some specific directions. Figure 2(b, c) show the real and imaginary parts of the band structure at a fixed $\Delta_\omega = -2.1$. As discussed above, band 1 is pure real and its imaginary part is not plotted in Fig. 2(c) for clearance. For $\alpha < \alpha_1 = 4.18$, the eigenvalues of band 3 and band 4 are complex conjugate pairs and that of band 2 is real; for $\alpha > \alpha_2 = 4.23$, the eigenvalues of band 2 and band 3 are complex conjugate while band 4 is real. $\alpha_1 < \alpha < \alpha_2$ denotes the corresponding bistable region where all three bands, i.e., bands 2-4, are real. Hence, the eigenspectra show similar behaviors as an EP$_2$ in a linear system at $\alpha = \alpha_1$ and $\alpha = \alpha_2$. We have also checked the critical behavior of the eigenvalue when approaching these two points (see Fig. S4), and the critical behaviors are also the same as an EP$_2$. For clarity, we denote these two points ELP$_2$ in the following text. When we vary $\Delta_\omega$, these two points form two arcs as highlighted by the solid yellow lines in Fig. 2(a). With the increase of $\Delta_\omega$, these two ELP$_2$ arcs gradually approach each other and merge when $\Delta_\omega = -2$ and $\alpha = \alpha_{EX}$ (denoted as an ELP$_3$). The corresponding complex band structures are shown in Figs. 2(d, e). Here, two of the bands are complex conjugate, and one of the bands remains real for $\alpha \neq \alpha_{EX}$. The critical behavior when approaching the ELP$_3$ is the same as a third-order EP in linear systems (see Fig. S4).

**Loss of orthogonality and the Sylvester matrix.**

All the eigenstates are orthogonal to each other in the Hermitian system, while the orthogonality is completely lost at the EPs in the non-Hermitian systems. The loss of orthogonality between two eigenstates can be captured with [31,32]

$$\chi_{ij} = 1 - |\langle \Psi_i | \Psi_j \rangle|. \tag{5}$$

Here, only the right-eigenvectors are used in Eq. (5). Correspondingly, $\chi_{ij} = 1$ for Hermitian systems, $\chi_{ij} \neq 1$ in general for non-Hermitian systems, and $\chi_{ij} = 0$ at the EPs. Figure 2(f) shows all the $\chi_{ij}$ versus $\alpha$ at $\Delta_\omega = -2.1$ [same parameters as in Figs. 2(b, c)]. It is clear that $\chi_{34} = 0$ at $\alpha = \alpha_1$ and $\chi_{23} = 0$ at $\alpha = \alpha_2$ where the two ELP$_2$ are reached. Except for these two ELP$_2$, $\chi_{ij}$ do not vanish. $\chi_{ij}$ for the parameters considered in Figs. 2(d, e) are provided in Fig. S5, where we can see that $\chi_{23} = \chi_{34} = \chi_{24} = 0$ at the ELP$_3$. Besides the orthogonality function $\chi_{ij}$, the Sylvester matrix is typically used to locate the EPs in non-Hermitian systems [33,34]. The Sylvester matrix is constructed based on the coefficients of the characteristic polynomial $P(\omega)$ in Eq. (4) and its derivative $P'(\omega)$. An EP is identified when the determinant of the Sylvester matrix [denoted as DSyl($P,P'$)] equals 0. We have checked that DSyl($P,P'$)=0 is satisfied at and only at the ELPs in Fig. 2(a). (Details provided in SM Sec. II)

**Mapping to higher-dimensional linear systems**

To properly characterize the topology of the EP-like features, one needs to embed the nonlinear system into a higher-dimensional linear system [27–29]. Since our system possesses four eigenstates, the lowest dimension of a linear system that can embed this nonlinear Hamiltonian is four. To address the mapping required above, we adopt a sophisticated mathematical tool named 'Isospectral Transformation' [30]. Isospectral matrix reduction is a powerful technique frequently employed in physics to address complex networks (described by matrices) by distilling high-dimensional matrices down to lower-dimensional ones. This method allows researchers to focus on a subset of 'nodes' or components within a system that are of particular interest, effectively reducing the complexity of analyzing the system's behavior. During the dimension reduction process, the spectrum of the original matrix is kept, and some basic or fundamental properties of the network are maintained. For instance, such a tool can be used to identify latent symmetries of complex higher-dimensional networks [35–38].

In our work, we perform the reverse process of isospectral reduction, i.e., expanding a $2 \times 2$ nonlinear Hamiltonian to a $4 \times 4$ linear Hamiltonian. Different from previous works where the mappings are isomorphic and hence the linear Hamiltonian is unique [27–29], here we have an infinite number of linear Hamiltonians that can be mapped to the nonlinear system in Eq. (1) under isospectral transformation. Below, we give one such linear Hamiltonian,

$$H_L(\alpha, \Delta_\omega, \kappa) = \begin{pmatrix} \alpha + \Delta_\omega & \kappa & -\sqrt{\kappa\alpha/2} & -i\sqrt{\kappa\alpha/2} \\ \kappa & 0 & 0 & 0 \\ -\sqrt{\kappa\alpha/2} & 0 & 0 & i\kappa \\ -i\sqrt{\kappa\alpha/2} & 0 & i\kappa & 0 \end{pmatrix}. \quad (6)$$

Details on obtaining $H_L$ are provided in SM Sec. III. Here we notice $H_L$ is non-Hermitian, and we can prove that all the possible linear Hamiltonians under isospectral transformation are non-Hermitian (proof in SM Sec. IV). Since $H_L$ is obtained through isospectral transformation, $H_L$ exhibits the same eigenspectra as shown in Fig. 2(a). Moreover, the yellow lines in Fig. 2(a) are mapped to lines of EP2 of $H_L$, and their merging point corresponds to an EP3. (See the SM Sec. III.) We emphasize that although the specific form of $H_L$ is not unique, the characteristic polynomial $P(\Omega)$ presented in Eq. (4) [as well as its derivative $P'(\Omega)$] is shared by all these linear Hamiltonians under mapping. Hence, the Sylvester matrix is the same, and thus all the possible $H_L$ under mapping share the same EPs in the parameter space.

**Chiral state transfer and frequency comb generation**

Chiral state transfer in linear non-Hermitian systems has drawn intense attention recently [17–20]. Chiral state transfer is usually present when the evolution of the system Hamiltonian encircles the EPs in the parameter space [17–19]. Given the same initial state, the final state depends only on the direction of encirclement, either clockwise (CW) or counterclockwise (CCW). Chiral state transfer also appears when circling an ELP in the parameter space. The mapping above builds a correspondence between a nonlinear Hamiltonian and a linear Hamiltonian. The dynamics governed by these two Hamiltonians are similar when the eigenfrequency is real. However, when the eigenfrequency is complex, the amplitude of the state increases or decreases exponentially in time for a linear Hamiltonian. In contrast, when considering the nonlinear Hamiltonian, the complex eigenfrequency is unphysical as the evolution of the state will quickly deviate from the requirement in Eq. (2). (The eigenstates being unphysical does not change the analysis of the ELPs above since the existence and order of EPs only concern the eigenspace, not the dynamics.) In other words, chiral state transfer may exhibit unique features that are different from their counterparts in linear systems.

Figure 3(a) plots the band structure for circling an $ELP_3$ (denoted by the yellow star) in the parameter space expanded by $\alpha$ and $\kappa$. The circling loop is centered at the $ELP_3$ with a radius of 0.1. The system is initially set to stay at one of the eigenstates on band 2 as marked by the blue diamond. After a time period $T = 2^{20}$, the system parameters return to its original position. During this period, the system parameters vary slowly and uniformly along the circle either CW or CCW. If the parameter encirclement is CCW (red arrow), the state of the system follows the eigenstates adiabatically as the parameters change. Hence, at $t = T$, the state ends at the eigenstate on band 4 (marked by the red diamond). This picture can be checked with the Fourier spectra of the instantaneous states. Figure 3(b) shows the spectra of the initial state and final state.

Compared with the initial state, the final state for CCW encirclement peaks at a different frequency which matches the eigenfrequency of band 4 at that point in the parameter space. The situation becomes different when considering the CW encirclement. The instantaneous state of the system still follows the eigenstates until encountering the ELP arc (the yellow line at the back). Noted, up to now, the dynamics of the system are still similar to previous works considering saturable gain [23,28]. Across the ELP arc, the instantaneous state becomes a mixed state exhibiting a comb spectrum, distinctly different from the linear counterparts. Different from previous works [23,28], our system is pure nonlinear without restoring force. Hence, the frequency comb will preserve till the end state as shown with the purple line in Fig. 3(b).

**Summary and discussion**

We have shown that the ELPs in our nonlinear Hermitian system can be mapped to EPs in higher-dimensional linear systems through isospectral transformation. Unique features of chiral state transfer are present in the encirclement of an ELP in the parameter space and a frequency comb is generated during this process. The physics discussed in our work is general and can be easily implemented on various platforms such as circuits [39], waveguides [40], optical ring resonators [41], acoustic cavities [42], and others [43]. We consider Kerr nonlinearity in our work, and other nonlinear effects should also be able to exhibit similar physics. Our work points to a new possibility of investigating intriguing phenomena associated with non-Hermitian terms in the pure Hermitian systems.

**Acknowledgment.**

Liang Fang and Kai Bai contribute equally to this work. The authors want to thank Prof. C. T. Chan, Prof. Z. Q. Zhang and Dr. Ruoyang Zhang for helpful discussion. This work is supported by the National Key Research and Development Program of China (Grant No. 2022YFA1404900), the National Natural Science Foundation of China (No.

12334015, No. 12274332, No. 12321161645)**.**

**Figures**

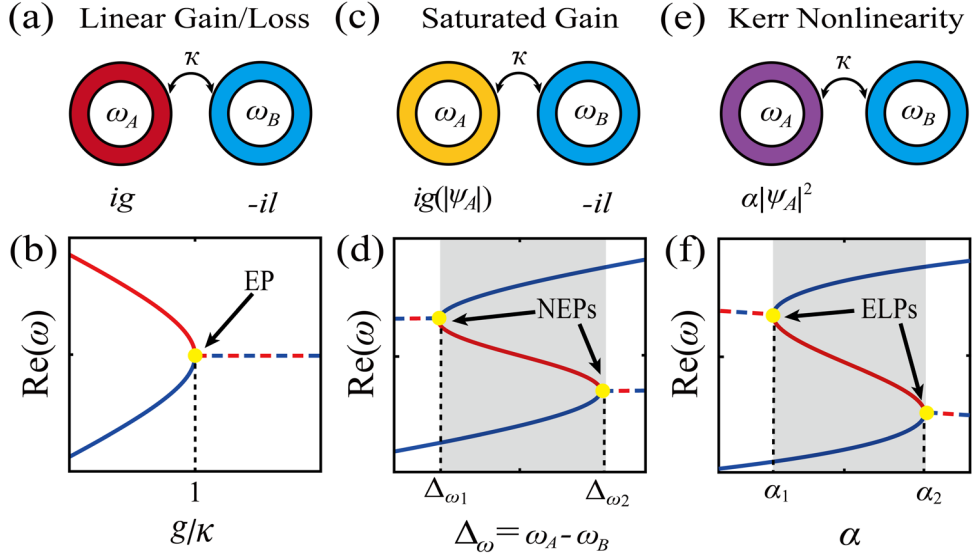

FIG. 1. (a) Two coupled resonators with linear gain and loss. When $\omega_A = \omega_B$ and $g = l$, such a system exhibits PT symmetry and possesses an exceptional point (EP) at $g/\kappa = 1$. (b) A sketch of the typical evolution of real parts of the eigenvalues in (a). (c) Two coupled resonators with a saturable gain $g(|\psi_A|)$ and linear loss. Such a nonlinear non-Hermitian system exhibits nonlinear EPs (NEPs). (d) The typical S-shape band structure of the system in (c) as one varies the detuning $\Delta_\omega$. NEPs appear at the turning points of the S-shape curve as denoted by the yellow dots. (e) Two coupled resonators where the resonance frequency of the left resonator depends on the wave amplitude through a nonlinear Kerr term $\alpha|\psi_A|^2$. (f) A sketch of the real part of the eigenvalues in (e) as one varies the Kerr coefficient $\alpha$. Kerr nonlinearity introduces a bi-stable region and the band structure behaviors as a S-shape curve. The turning points (marked by the yellow dots) of this curve are denoted as EP-like points (ELPs).

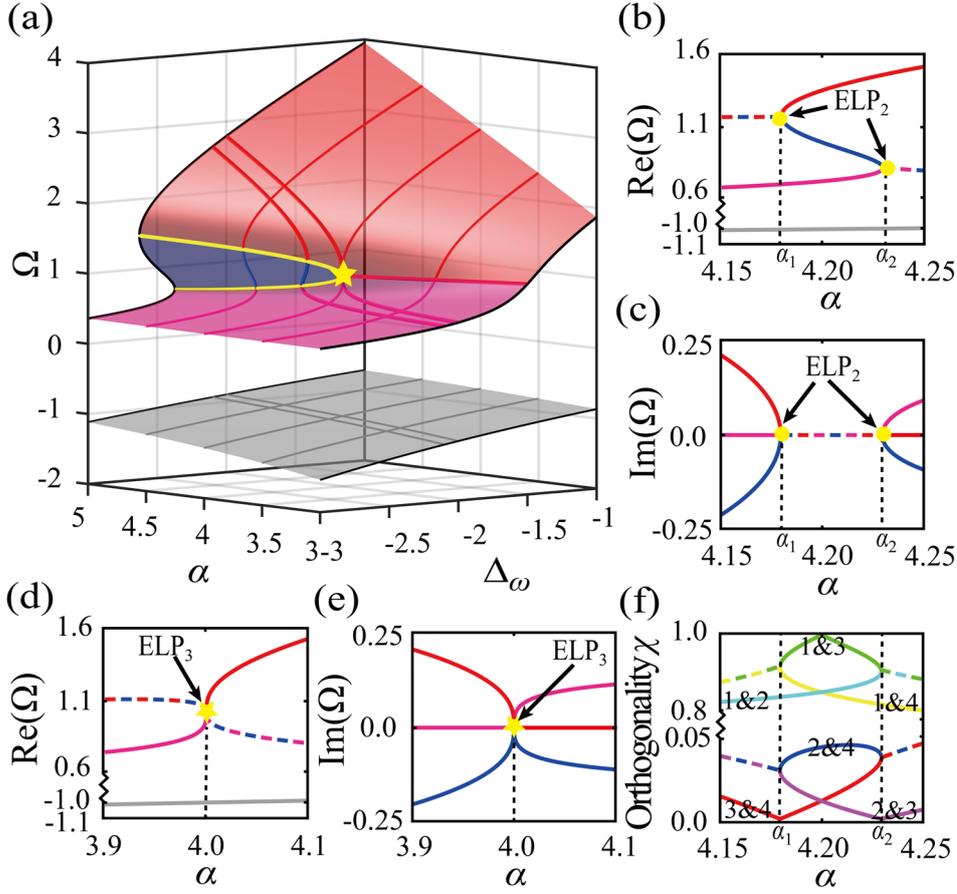

FIG. 2. (a) Eigenspectra of the nonlinear Hamiltonian $H_{NL}$ versus the detuning $\Delta_\omega$ and the Kerr coefficient $\alpha$. Here only the eigenmodes with pure real eigenfrequency are shown. The light gray, pink and red surfaces represent stable modes, and the blue surface denotes unstable modes. The yellow lines represent the arcs of ELPs and these two arcs merge at a higher-order ELP as marked by the yellow star. (b, c) Real (b) and imaginary (c) parts of the eigenfrequency with a fixed detuning $\Delta_\omega = -2.1$ where two ELP$_2$ (yellow dots) can be seen. (d, e) Real (d) and imaginary (e) parts of the eigenfrequency at $\Delta_\omega = -2$ and an ELP$_3$ is observed at $\alpha_{EX} = 4$. (f) $\chi_{ij}$ versus $\alpha$ at $\Delta_\omega = -2.1$. In (b, c, f), $\alpha_1 = 4.18$ and $\alpha_2 = 4.23$, and $\kappa = 1$ in all the subfigures.

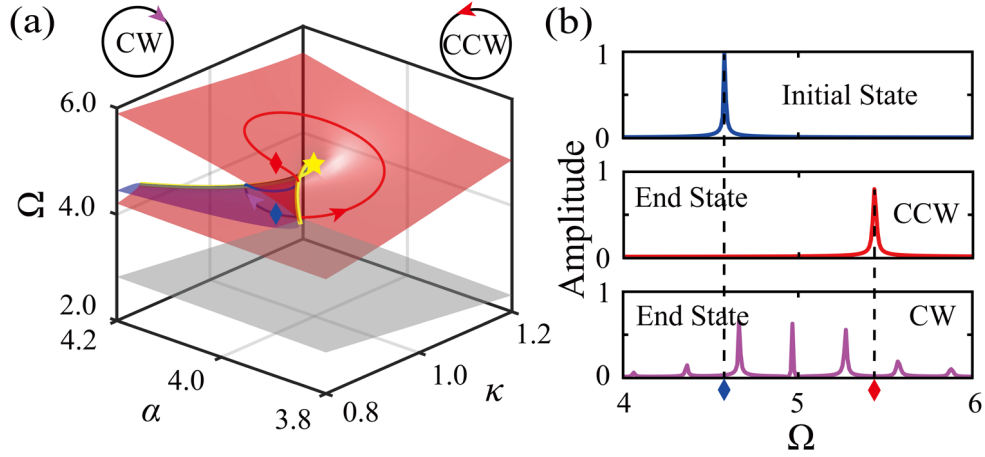

FIG. 3. (a) Spectra of the modes with real eigenfrequency versus the Kerr coefficient $\alpha$ and the coupling between two resonators $\kappa$. The circle indicates the route in the parameter space circling an ELP$_3$ (marked by the yellow star). The system is assumed to start with an eigenstate on band 2 inside the bistable region (as denoted by the blue diamond), and then evolve clockwise (CW, purple arrow) or counterclockwise (CCW, red arrow) and slowly enough along the route. (b) The Fourier spectra of the initial state (blue) and final states for CCW (red) and CW (purple) evolutions. Here the variations of parameters are given by $\alpha = \alpha_{EX} + 0.1\sin(\pm 2\pi t/T)$ and $\kappa = 1 - 0.1\cos(\pm 2\pi t/T)$ for $t \in [0, T]$, and $\pm$ chooses the evolution direction. The circle is divided into $N$ continuous segments, and we perform FFT for each segment to track the spectra of the state. Here, the blue curve is the FFT result of the first segment, and the red and purple curves are the FFT results of the last segment. When $N$ is large enough ($N = 64$ in our simulation), the spectra in (b) approximately represent the initial and final states. In the simulations, $\Delta_\omega = -2$ and $T = 2^{20}$.